\documentclass[10pt,a4paper]{article}
\usepackage[latin1]{inputenc}
\usepackage{amsmath}
\usepackage{amsfonts}
\usepackage{graphicx}
\usepackage{bm}
\input epsf
\usepackage[margin=4cm]{geometry}

\usepackage{hyperref} 
\hypersetup{
colorlinks=true,
allcolors=blue
}

\begin{document}

\title{Parallel axis theorem for free-space electron wavefunctions}

\author{Colin R Greenshields, Sonja Franke-Arnold and Robert L Stamps \\ \\
SUPA School of Physics and Astronomy, \\
University of Glasgow, Glasgow G12 8QQ, UK \\ \\
Email: Sonja.Franke-Arnold@glasgow.ac.uk, \\ Robert.Stamps@glasgow.ac.uk
}

\date{31 August 2015}

\maketitle

\begin{abstract}
We consider the orbital angular momentum of a free electron vortex moving in a uniform magnetic field.  We identify three contributions to this angular momentum: the canonical orbital angular momentum associated with the vortex, the angular momentum of the cyclotron orbit of the wavefunction, and a diamagnetic angular momentum.  The cyclotron and diamagnetic angular momenta are found to be separable according to the parallel axis theorem.  This means that rotations can occur with respect to two or more axes simultaneously, which can be observed with superpositions of vortex states.
\end{abstract}


\section{Introduction}

In classical mechanics, an electron moving in a uniform magnetic field follows a circular orbit in the plane perpendicular to the field, as dictated by the Lorentz force.  This so-called cyclotron motion, which occurs even in the absence of a central potential, plays a role in areas as diverse as particle physics \cite{Lawrence1932, Veksler1945}, electron microscopy \cite{Hawkes1996}, plasma physics \cite{Erckmann1994}, and also in microwave ovens \cite{Vollmer2004}.

Electrons can also possess quantized canonical orbital angular momentum which does not depend on the presence of a magnetic field.  This is well known in the case of bound states in atoms and quantum dots \cite{Millo2001}, in which there is a confining potential, however recently it was discovered that \emph{free} electrons can be imprinted with orbital angular momentum.  Electron vortex beams \cite{Bliokh2007, Uchida2010, Verbeeck2010, McMorran2011}, generated, for example, in electron microscopes, have twisted wavefronts, and resemble freely propagating atomic orbitals.  The understanding, manipulation and exploitation of this angular momentum for a range of technological applications is currently a very active area of investigation \cite{Bliokh2011, Karimi2012, Lloyd2012a, Lloyd2012b, Hemsing2013, Ivanov2013, Beche2014, AsenjoGarcia2014, Hayrapetyan2014, Matula2014, Grillo2015}.

In a magnetic field, an electron can possess both canonical orbital angular momentum, \emph{and} angular momentum arising from the interaction with the field.  If the magnetic field is uniform, the canonical angular momentum in the direction of the field is independent of the field and is constant \cite{Gallatin2012}.  Meanwhile, the magnetic field induces an additional current within the electron's wavefunction which gives rise to a diamagnetic angular momentum \cite{Landau1930, Darwin1931, Greenshields2014, Babiker2015}.  Manipulating the canonical and diamagnetic orbital angular momenta of free electrons recently led to the first direct imaging of Landau states \cite{Schattschneider2014, Schachinger2015}.  In addition to the diamagnetic rotation of an electron's wavefunction, however, in a magnetic field there will generally also be a cyclotron orbit of the centre of mass of the wavefunction \cite{Gallatin2012, Bliokh2012a}.  The angular momentum associated with this cyclotron motion has not previously been considered.

In this paper, we show that the total orbital angular momentum of the electron is described by the parallel axis theorem.  This angular momentum comprises the canonical and diamagnetic components, which are associated with rotation relative to the centre of mass of the wavefunction, and a cyclotron component which has expectation value equal to that for the classical orbit.  Interestingly, for free electrons all three of these components can have similar magnitude.  This means that the trajectory of the electron is strongly dependent on how these angular momenta add and subtract.  Further, we show that different cyclotron orbits can be superposed, leading to rotations with respect to multiple parallel axes, and periodic interference.  Our results suggest novel means of structuring electron beams for use in specific applications, such as probing magnetic and chiral materials.

\section{Model}

We consider an otherwise free electron moving under the influence of a uniform magnetic field.  We take the direction of this magnetic field to define our $z$ axis, and consider the motion of the electron within the $x$-$y$ plane.  The electron may also be moving in the $z$ direction, however the component of its momentum in this direction is a constant of motion \cite{Greenshields2014}, and will not affect our results.  We consider non-relativistic energies, meaning that the spin angular momentum is also constant, and can be separated from the orbital motion of the electron \cite{Greenshields2012}.  In what follows, we shall consider only the electron's orbital angular momentum.  The magnetic field $\bm{B}=B\hat{\bm{z}}$ can be described by the cylindrically symmetric vector potential $\bm{A}=B\rho\bm{\hat{\phi}}/2$, where $\rho$ and $\phi$ are cylindrical polar coordinates.  This choice of gauge is convenient as it allows us to exploit the rotational symmetry of the magnetic field.  Our Hamiltonian is therefore
\begin{equation}
\label{H}
H = \frac{(\bm{p}_\perp^{\rm kin})^2}{2m} = \frac{1}{2m}\left[(\bm{p}_\perp^{\rm can})^2+\frac{1}{4}e^2B^2\rho^2-eBL^{\rm can}_z\right],
\end{equation}
where $\bm{p}^{\rm kin}_\perp=m\bm{v}_\perp=\bm{p}_\perp^{\rm can}-e\bm{A}_\perp$ is the component of the electron's kinetic momentum in the plane perpendicular to the magnetic field, $\bm{p}_\perp^{\rm can}=-i\hbar\bm{\nabla}_\perp$ is the corresponding canonical momentum component, $L^{\rm can}_z = \rho p^{\rm can}_\phi = -i\hbar\,\partial/\partial\phi$
is the $z$-component of the canonical orbital angular momentum,
$e=-|e|$ is the electron's charge, and $m$ its mass.

We are interested in the evolution of non-stationary states of the system, described by the time-dependent Schr\"{o}dinger equation
\begin{equation}
\label{SE}
{\rm i}\hbar\frac{\partial \Psi(\bm{r}_\perp,t)}{\partial t}=H\Psi(\bm{r}_\perp,t),
\end{equation}
where $\bm{r}_\perp=(\rho,\phi)$ is the position of the electron in the plane perpendicular to the magnetic field.  Note that if the electron is moving along the $z$ axis with a velocity $v_z$, the Schr\"{o}dinger equation \eqref{SE} describes the state of the electron after a propagation distance of $z=v_zt$ \cite{Schattschneider2014}.

An electron with momentum transverse to the magnetic field will exhibit cyclotron motion.  This is conventionally described in a classical context.  Here we will derive the cyclotron motion by assuming an electron wavefunction
\begin{equation}
\label{Psi0}
\Psi(t=0)=\Psi_0=u(\rho)\exp\left[{\rm i}\left(\ell\phi+\frac{p_{\rm c}}{\hbar}x\right)\right],
\end{equation}
where $\ell\in\mathbb{Z}$.  We have defined the $x$ axis as the direction of the transverse kinetic momentum at $t=0$.  This state has a rotationally symmetric probability density $|\Psi_0|^2=|u(\rho)|^2$, and an expectation value of canonical angular momentum $\langle L_z^{\rm can}\rangle=\ell\hbar$.  We shall see that the momentum $\langle\bm{p}_\perp^{\rm kin}\rangle_0=p_{\rm c}\hat{\bm{x}}$ results in a cyclotron orbit of the wavefunction.

\begin{figure}
\centering
\includegraphics[width=8.64cm]{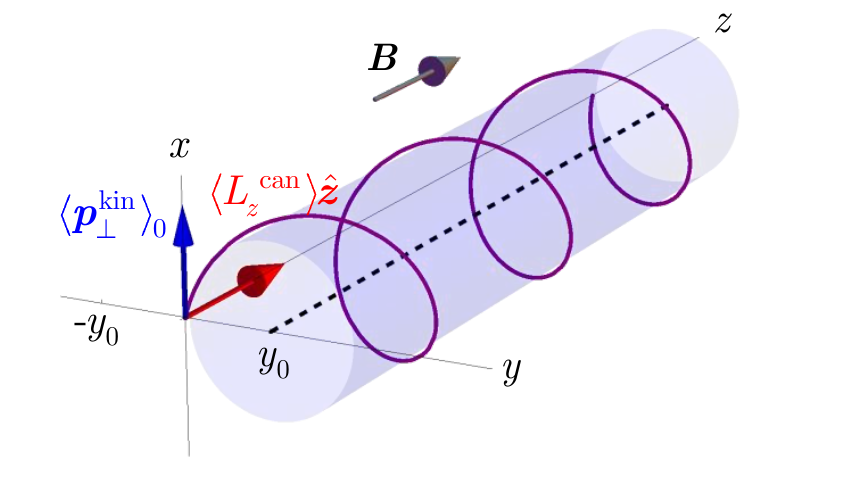}
\caption{Cyclotron trajectory of the centre of mass of the wavefunction.  This orbit occurs with respect to the axis $y=y_0$, the position of which depends on the initial transverse momentum $\langle\bm{p}_\perp^{\rm kin}\rangle_0$ as well as the magnetic field.  Also indicated is the direction of the canonical orbital angular momentum $\langle L^{\rm can}_z\rangle\hat{\bm{z}}$.}
\label{Fig1}
\end{figure}

Note that in our model the canonical orbital angular momentum is not collinear with the instantaneous direction of propagation of the wavefunction.  This is illustrated in figure~\ref{Fig1}.  The angular momentum is in the direction of the magnetic field, while the kinetic momentum has a component perpendicular to the magnetic field.  This contrasts with vortex states, either in field-free space or in a magnetic field, which are energy eigenfunctions, as these have momentum and angular momentum which are collinear \cite{Bliokh2012a, Lloyd2013}.  Electrons in non-stationary states can have angular momentum at an arbitrary angle to their direction of propagation, however \cite{Gallatin2012, Bliokh2012b, Bialynicki2000}.  Here, as a result of the cyclotron orbit, the \emph{time-averaged} expectation value of kinetic momentum is collinear with the angular momentum $\langle L^{\rm can}_z\rangle$.

\section{Electron trajectories and angular momentum}

In the following we will show that the different forms of angular momentum give rise to rotations with respect to more than one axis.  This can be seen by examining the ``trajectories'' associated with the electron's probability distribution and current density.

First, we will consider the expectation value of the electron's position, which is equivalent to the centre of mass of its probability distribution.
Differentiating twice with respect to time, we obtain the equation of motion
\begin{eqnarray}
\label{D2r}
\frac{\partial^2\langle \bm{r}_\perp\rangle(t)}{\partial t^2} &=&
-\frac{1}{\hbar^2}\left\langle\left[[\bm{r}_\perp,H],H\right]\right\rangle(t) \nonumber \\
& = & -\omega_{\rm c}^2\left\langle\bm{r}_\perp-\left[x+p_y^{\rm kin}/(eB)\right]\hat{\bm{x}}-\left[y-p_x^{\rm kin}/(eB)\right]\hat{\bm{y}}\right\rangle(t) \nonumber \\
& = &  -\omega_{\rm c}^2\left(\langle\bm{r}_\perp\rangle(t)-y_0\hat{\bm{y}}\right),
\end{eqnarray}
where $\omega_{\rm c}=-eB/m$ is the cyclotron angular velocity and $y_0=p_{\rm c}/(|e|B)$.  Here we have used the fact that the quantity
$\left[x+p_y^{\rm kin}/(eB)\right]\hat{\bm{x}}+\left[y-p_x^{\rm kin}/(eB)\right]\hat{\bm{y}}$,
which is the centre of the orbit of a classical particle which has position $\bm{r}_\perp$ and momentum $\bm{p}_\perp^{\rm kin}$ \cite{Li1999}, has the constant expectation value $y_0\hat{\bm{y}}$. The initial position and velocity of the centre of mass of the probability distribution are given by $\langle\bm{r}_\perp\rangle(0)=\mathbf{0}$ and
\begin{equation}
\frac{\partial\langle \bm{r}_\perp\rangle(0)}{\partial t} =
-\frac{{\rm i}}{\hbar}\left\langle[\bm{r}_\perp,H]\right\rangle(0)
=\frac{p_{\rm c}}{m}\hat{\bm{x}}
\end{equation}
respectively, and substituting these into \eqref{D2r} yields the trajectory
\begin{equation}
\label{r}
\langle \bm{r}_\perp\rangle(t)=y_0[\sin \omega_{\rm c}t\hat{\bm{x}}+ (1-\cos\omega_{\rm c}t)\hat{\bm{y}}].
\end{equation}
This trajectory, illustrated in figure~\ref{Fig1}, is a circular orbit with radius
\begin{equation}
\sigma=|y_0|=\left|\frac{p_{\rm c}}{eB}\right|
\label{sigma}
\end{equation}
and angular velocity $\omega_{\rm c}$ -- the cyclotron orbit of a classical particle with the momentum $p_{\rm c}$.  The trajectory of the centre of mass of the probability distribution is therefore independent of the canonical angular momentum of the electron.

The canonical angular momentum is instead associated with a circulation of current within the electron's probability distribution.  This can be seen by examining the probability current density $\bm{j}_\perp(\bm{r}_\perp,t)={\rm Re}(\Psi^*\bm{p}_\perp^{\rm kin}\Psi)/m$.  To do so we have solved the time-dependent Schr\"{o}dinger equation \eqref{SE} numerically using the Chebyshev method \cite{TalEzer1984, vanDijk2011, Dziubak2012}, as described in Appendix~\ref{Appendix}.  We must first specify the radial distribution, $u(\rho)$, of the initial wavefunction $\eqref{Psi0}$.  Here we will set this to be the same as that of a Landau state -- one of the energy eigenstates of the system:
\begin{equation}
\label{Lan}
u(\rho)=u_{n,|\ell|}^{\rm Lan}(\rho)=N_{n,|\ell|}\left( \frac{\rho\sqrt{2}}{\rho_B} \right)^{|\ell|}\exp \left( -\frac{\rho^2}{\rho_B^2} \right) L_n^{|\ell|}\left( \frac{2\rho^2}{\rho_B^2} \right),
\end{equation}
where $n=0,1,2,...$ is the radial quantum number, $L_n^{|\ell|}$ is an associated Laguerre polynomial, $\rho_B=\sqrt{4\hbar/|eB|}$ is the width of the Gaussian envelope and $N_{n,|\ell|}=\sqrt{2n!/[\pi(n+|\ell|)!]}/\rho_B$ is a normalisation constant.  This means that if $p_{\rm c}=0$, the electron would be in a Landau state.  An arbitrary radial distribution could be decomposed in terms of the eigenfunctions $u_{n,|\ell|}^{\rm Lan}$.

\begin{figure}
\centering
\includegraphics[width=10cm]{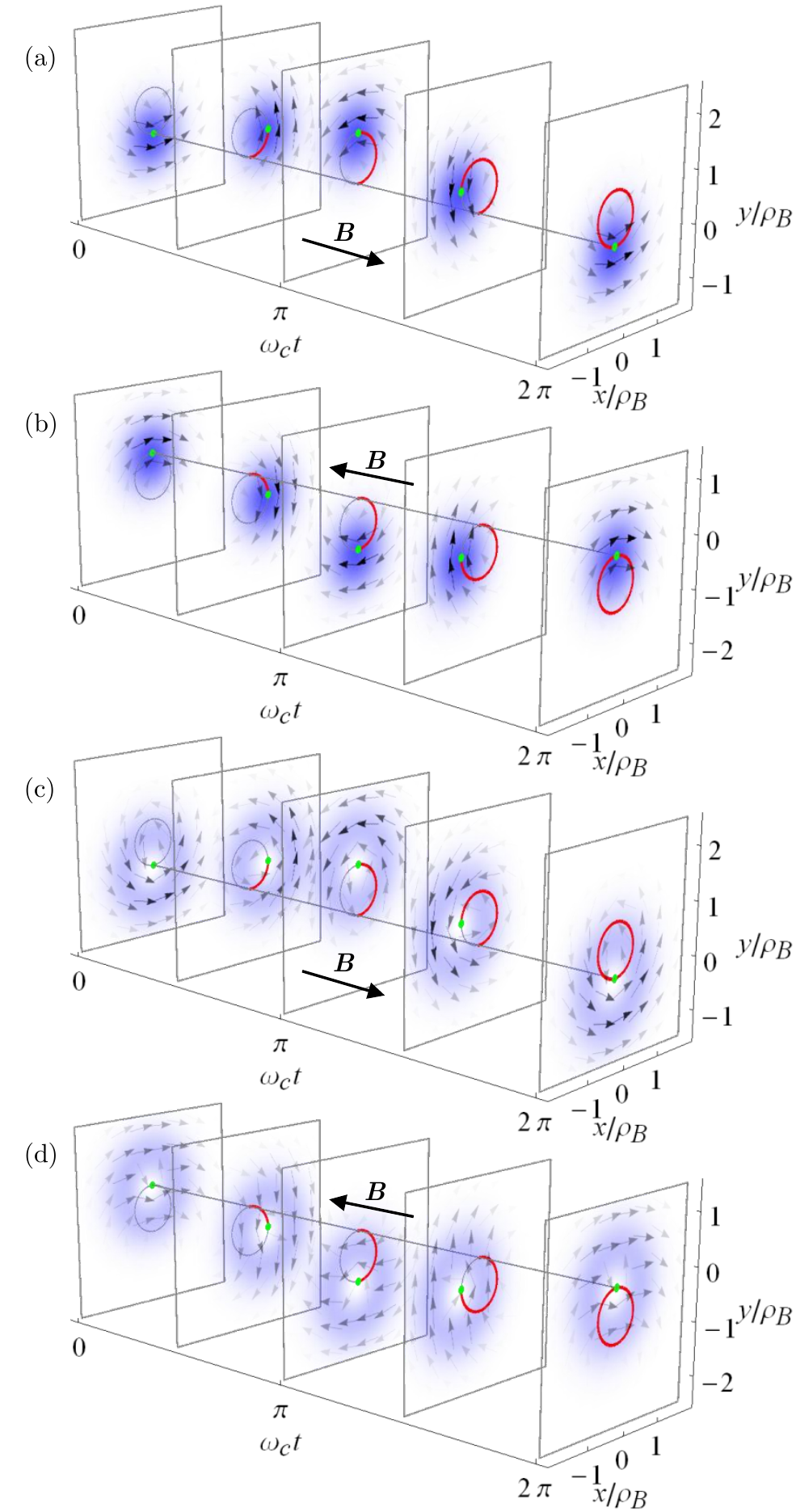}
\caption{Probability density $|\Psi|^2$ and current density $\bm{j}_\perp$ for wavefunctions with canonical orbital angular momentum $\ell=0$ (a, b) and $\ell=1$ (c, d), for opposite directions of the magnetic field.  In each case the transverse momentum is $p_{\rm c}=2\hbar/\rho_B$ and the wavefunction has the radial distribution $u_{0,|\ell|}^{\rm Lan}$.  The red arcs indicate the trajectory of the centre of mass of the probability distribution, which is highlighted in green.}
\label{Fig2}
\end{figure}

The time-evolution of the probability density $|\Psi|^2$ and the current density $\bm{j}_\perp$ are shown in figure~\ref{Fig2}.  Here the transverse momentum $p_{\rm c}$ has been chosen such that the radius of the cyclotron orbit is approximately equal to the width of the probability distribution.  In (a) and (b) the electron has no net canonical orbital angular momentum, while in (c) and (d) it has a canonical orbital angular momentum $\ell=1$.  The evolution of these states is shown for different directions of the magnetic field, which result in different directions of the cyclotron orbit. Whereas the probability density follows a straightforward classical orbit, the current density is seen to depend in a non-trivial manner on both the wavefunction and the magnetic field. In particular, in contrast to the classical cyclotron trajectory, and also to orbital angular momentum eigenstates in the absence of a magnetic field, the current distribution here is not rotationally symmetric. The rotational symmetry of the probability distribution, with respect to its centre of mass, is preserved, however. This reflects the fact that the magnetic field is rotationally symmetric, and the canonical angular momentum $L_z^{\rm can}$ is conserved.  This means that the canonical orbital angular momentum of the electron is associated with a rotation axis at the centre of mass of the probability distribution, and is independent of the cyclotron orbit.

\begin{figure}
\centering
\includegraphics[width=9.5cm]{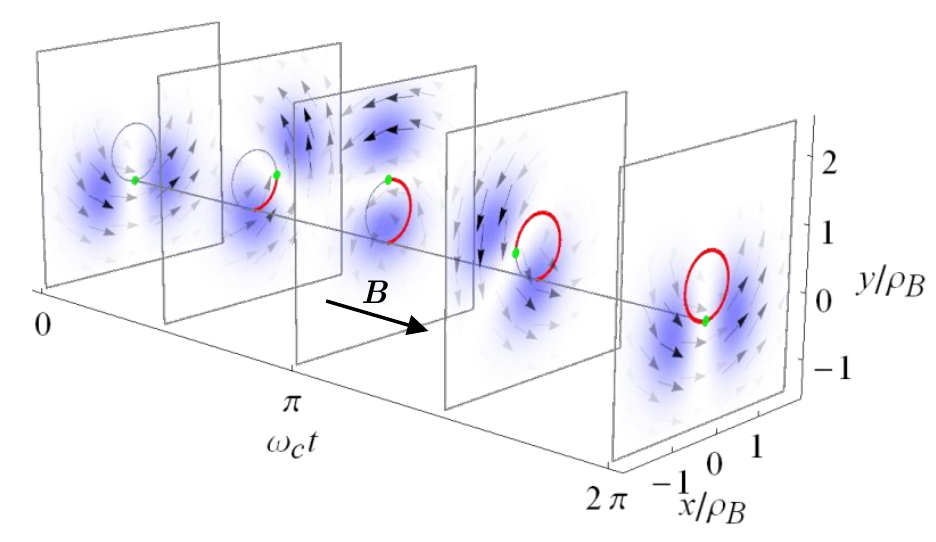}
\caption{Probability density and current density for an equally weighted superposition of wavefunctions which both have transverse momentum $p_{\rm c}=2\hbar/\rho_B$, but have opposite values of canonical angular momentum $\ell=\pm1$.  The constituent states have the same radial distributions as in figure~\ref{Fig2}.}
\label{Fig3}
\end{figure}

More surprising, perhaps, is that rotations in fact occur with respect to the cyclotron axis \emph{and} the centre of mass axis even when the electron does not possess any net canonical angular momentum.  This is as a result of the diamagnetic angular momentum that any electron wavefunction possesses in the presence of a magnetic field \cite{Landau1930, Darwin1931}.  This angular momentum arises as a result of the circulating current the magnetic field induces within the wavefunction, and is associated with a rotation of the probability density at the Larmor angular velocity $\omega_{\rm L}=\omega_{\rm c}/2=-eB/(2m)$ \cite{Schattschneider2014, Guzzinati2013}.  The diamagnetic angular momentum is equal to $L_z^{\rm dia}=I'\omega_{\rm L}$, where
\begin{equation}
\label{Iprime}
I'=m\langle\rho'^2\rangle
\end{equation}
is the moment of inertia of the electron's probability distribution, in the reference frame of its centre of mass \cite{Greenshields2014}.  Here $\rho'=|\bm{r}_\perp-\langle\bm{r}_\perp\rangle|$ is the radial coordinate in this reference frame. This angular momentum has the same direction as the external magnetic field, meaning that the associated magnetic moment, $\mu_z^{\rm dia}=eL_z^{\rm dia}/(2m)$, opposes the external field, in accordance with Lenz's law. In contrast to cyclotron motion, the diamagnetic rotation depends on there being an extended probability distribution, and vanishes in the classical limit. For the Landau radial distribution $u_{n,\ell}^{\rm Lan}$, the mean square radius is $\langle\rho'^2\rangle_{n,\ell}^{\rm Lan}=(2n+|\ell|+1)\rho_B^2/2$, and the diamagnetic angular momentum therefore takes the quantized values $L_z^{\rm dia,\, Lan}={\rm sign}(B)(2n+|\ell|+1)\hbar$ \cite{Bliokh2012a}. The effect of the diamagnetic angular momentum becomes clear when we consider a superposition of opposite values of canonical orbital angular momentum, such as that shown in figure~\ref{Fig3}.  As this superposition has no net canonical angular momentum, the rotation of the electron's probability density with respect to its centre of mass is due entirely to the diamagnetic angular momentum. Previously we have interpreted this as a form of Faraday rotation for electrons \cite{Greenshields2012}.

In general, the orbital angular momentum with respect to the centre of mass axis will be given by a sum of canonical and diamagnetic contributions.  The motion is thus described by two independent rotations: the cyclotron orbit, and the rotation around the instantaneous centre of mass axis due to the canonical and diamagnetic angular momenta.

\section{Parallel axis theorem}

The rotation of the electron's wavefunction is reminiscent of a classical rigid body.  We will explore this analogy further by considering the kinetic angular momentum of the electron, which is the total mechanical angular momentum it possesses while moving in the magnetic field.  This angular momentum has the $z$ component
\begin{equation}
L_z^{\rm kin}=\rho p_\phi^{\rm kin}=L_z^{\rm can}-\frac{1}{2}eB\rho^2,
\label{Lkin}
\end{equation}
with an expectation value of
\begin{equation}
\label{LkinExp}
\langle L_z^{\rm kin}\rangle=\ell\hbar+I\omega_{\rm L}
\end{equation}
for any state with a canonical orbital angular momentum $\ell\hbar$.
Here $I = m\langle\rho^2\rangle$ is the moment of inertia of the electron's probability distribution for rotation with respect to the $z$ axis.  Just as with a rigid body, we can use the parallel axis theorem to express the moment of inertia $I$ as a sum of two components:
\begin{equation}
\label{I}
I=m\rho_0^2+I',
\end{equation}
where $\rho_0=|\langle\bm{r}_\perp\rangle|=\sqrt{2(1-\cos\omega_{\rm c}t)}\,\sigma$, with $\sigma$ defined by \eqref{sigma}, is the radial coordinate of the centre of mass, and $I'$ is the moment of inertia with respect to the centre of mass axis, given by \eqref{Iprime}.  These two components correspond to the cyclotron orbit of the wavefunction and its diamagnetic angular momentum respectively. The total kinetic angular momentum of the electron, which we obtain from \eqref{LkinExp} and \eqref{I}, can therefore be expressed as
\begin{equation}
\label{LkinExp2}
\langle L_z^{\rm kin}\rangle=\ell\hbar+L_z^{\rm cyclo}+L_z^{\rm dia},
\end{equation}
where $L_z^{\rm cyclo} = m\omega_{\rm L}\rho_0^2=(1-\cos\omega_{\rm c}t)m\omega_{\rm c}\sigma^2$ is the angular momentum associated with the cyclotron orbit. While the relation between the kinetic and canonical angular momenta in \eqref{Lkin} is true also for a classical point particle \cite{Reimer2008}, the decomposition into separate cyclotron and diamagnetic components which follows from \eqref{I} is only meaningful for an extended probability distribution.

Unlike the canonical and diamagnetic components, the cyclotron angular momentum depends on our choice of reference axis.  A natural choice is to consider the angular momentum with respect to the centre of the cyclotron motion.  In this reference frame, which we reach by making the transformation $y\rightarrow\tilde{y}=y-y_0$, the cyclotron angular momentum has the constant value
\begin{equation}
\tilde{L}_z^{\rm cyclo}=m\omega_{\rm c}\sigma^2.
\end{equation}
Irrespective of the reference frame, of course, the total kinetic angular momentum will be equal to the sum of the three components described.

One may expect that the cyclotron angular momentum, which exists classically, would be the dominant contribution to the electron's kinetic angular momentum.  However, this need not be the case, as can be seen by considering typical parameters for free electrons in electron microscopes.  For example, if an electron beam which is initially propagating parallel to a magnetic field is transmitted through a diffraction grating with a period $d=100\,{\rm nm}$, the first diffraction order will have a net transverse momentum of $p_{\rm c}=\hbar(2\pi/d)=h/d$, which corresponds to an energy of $p_{\rm c}^2/(2m)=h^2/(2md^2)=0.15\,{\rm meV}$. In a magnetic field of $B=1\,{\rm T}$, the resulting cyclotron orbit, which has radius $41\,{\rm nm}$, will have an angular momentum of $2.6\hbar$.  This is of the same order of magnitude as the canonical angular momentum of the lowest order vortex states, and considerably smaller than that of vortex beams recently generated with a winding number of $\ell=200$ \cite{Grillo2015}.  Indeed, in a given magnetic field, the cyclotron angular momentum can in principle have any size, ranging from zero to macroscopic values, depending on the net transverse momentum $p_{\rm c}$.  The diamagnetic angular momentum can also take a wide range of values, as the mean square radius of the probability distribution is varied \cite{Greenshields2014}, although this has a lower limit due to the uncertainty principle and a maximum due to the requirement of spatial coherence.  This means that the different rotations we have described can indeed occur on the same length scale, justifying the choices of parameters in our figures.

\section{Superposition of cyclotron orbits}

\begin{figure}
\centering
\includegraphics[width=9.5cm]{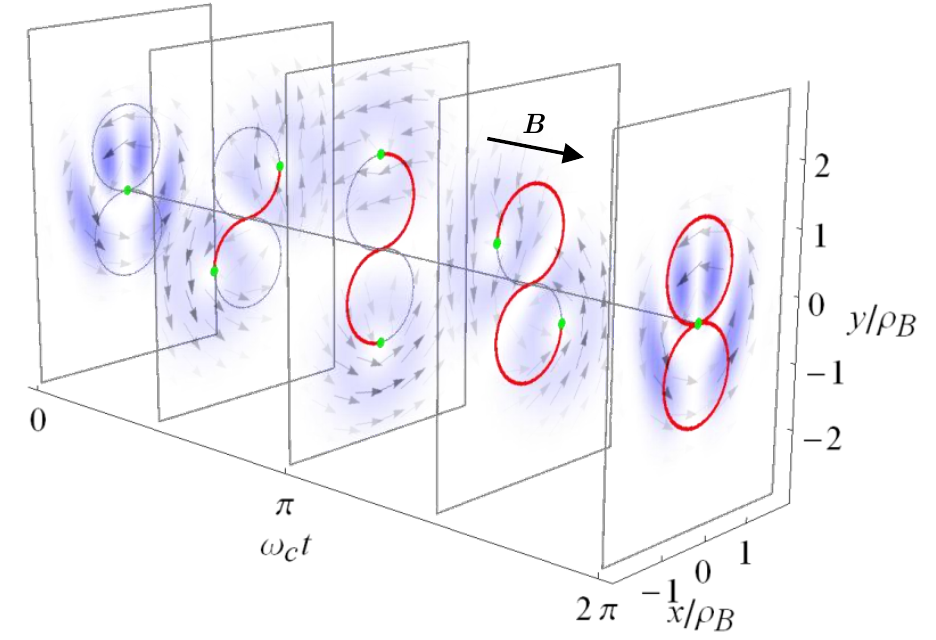}
\caption{Probability density and current density for an equally weighted superposition of a wavefunction with $p_{\rm c}=3\hbar/\rho_B$ and $\ell=-1$ and a wavefunction with $p_{\rm c}=-3\hbar/\rho_B$ and $\ell=1$.  The constituent states have the same radial distributions as in figures~\ref{Fig2} and~\ref{Fig3}.  The centre of mass of the superposition remains stationary in the $x$-$y$ plane, while the individual components follow the cyclotron trajectories indicated.}
\label{Fig4}
\end{figure}

So far we have considered rotations with respect to two different axes -- the cyclotron axis, as well as the centre of mass axis.  The position of the cyclotron axis was defined by the transverse momentum $p_{\rm c}$ which appears in a plane wave factor in the wavefunction \eqref{Psi0}.  Suppose, however, that we have a superposition of different transverse momenta. Arbitrary distributions of transverse momentum could be created using appropriately designed holograms \cite{Verbeeck2014, VolochBloch2013, Grillo2014, Harvey2014}.  A simple example would be the following superposition of two momenta, $p_{\rm c,1}$ and $p_{\rm c,2}$, which may also be associated with different canonical orbital angular momenta, $\ell_1$ and $\ell_2$:
\begin{equation}
\label{Psi0_super}
\Psi_0=\frac{u^{\rm Lan}_{n,|\ell_1|}(\rho)}{\sqrt{2}}\exp\left[{\rm i}\left(\ell_1\phi+\frac{p_{\rm c,1}}{\hbar}x\right)\right]+ \frac{u^{\rm Lan}_{n,|\ell_2|}(\rho)}{\sqrt{2}}\exp\left[{\rm i}\left(\ell_2\phi+\frac{p_{\rm c,2}}{\hbar}x\right)\right].
\end{equation}
This is similar to the wavefunction formed when a plane wave is transmitted through a forked diffraction grating \cite{McMorran2011}. For the discussion here the particular form of the radial distribution is not important, and the Landau function has been chosen with numerical efficiency in mind. The evolution of such a state, here with equal and opposite values of both the transverse momentum and canonical angular momentum, is shown in figure~\ref{Fig4}. It can be seen that there are now two cyclotron orbits, which have different rotation axes.  These are associated with different directions of the initial momentum $p_{\rm c}$.  As a result, the two components of the superposition move apart, before re-combining and interfering.

Taken together, the cyclotron orbits in figure~\ref{Fig4} describe a rotation, with respect to the $z$ axis, at the Larmor angular velocity $\omega_{\rm L}$. This is consistent with the predictions of classical electron optics regarding image formation in rotationally symmetric magnetic lenses \cite{Hawkes1996}. Interestingly, though, in our case the axis of the Larmor rotation is not defined by a symmetry of the magnetic field -- a uniform magnetic field is rotationally symmetric with respect to an infinite number of axes. Rather, here the Larmor rotation occurs with respect to the centre of mass of the electron's probability distribution. This is the case both in figure~\ref{Fig3}, where the centre of mass follows a cyclotron orbit, and in figure~\ref{Fig4}, where the centre of mass is stationary. Further, it must be remembered that we are considering here a single electron which is in a state of superposition. This means, for example, that if one of the two cyclotron components in figure~\ref{Fig4} underwent an interaction which modified its phase, this could be detected through its effect on the subsequent interference pattern.

\section{Summary and outlook}

In summary, we have shown that in a magnetic field an electron can rotate around more than one axis simultaneously.  The wavefunction of the electron follows a cyclotron orbit, and superposed onto this is a rotation around the instantaneous centre of mass.  The rotation with respect to the centre of mass axis arises as a result of the diamagnetic angular momentum, as well as any canonical orbital angular momentum the electron possesses.  The kinetic angular momentum of the electron is therefore described by the parallel axis theorem.

Our results show that canonical orbital angular momentum and cyclotron motion provide separate degrees of freedom for shaping electron current distributions.  This could allow electron beams to be structured for use in specific applications.  For example, the symmetry of the current distribution could be optimized to probe specific transitions in materials \cite{Verbeeck2014, Rusz2014}. It may also be possible to utilise cyclotron trajectories in novel forms of interferometry. Moreover, here we have only considered the case in which the canonical angular momentum and magnetic field are parallel, so that the rotation is confined to a plane.  With canonical angular momentum and magnetic fields which are in different directions to one another, the angular momentum and current density could be shaped in three dimensions.

Further, if the angular momenta were in different directions, it appears that they would become coupled.  Canonical orbital angular momentum which is at an angle to a uniform magnetic field would be expected to precess around the direction of the field \cite{Gallatin2012, Bialynicki2000}.  The canonical angular momentum is also not conserved when the rotational symmetry of the magnetic field is broken, such as in astigmatic magnetic lenses \cite{Schattschneider2012, Clark2013, Peterson2013}.  Not only this, but in non-uniform magnetic fields the spin and orbital degrees of freedom of an electron with non-relativistic velocity are no longer independent \cite{Karimi2012, Gallup2001}.  The nature of the coupling between all of these angular momenta is an interesting avenue for future investigation.

\section*{Acknowledgements}

C. G. is supported by a SUPA Prize Studentship.  We also acknowledge support from the UK EPSRC under Grant No. EP/M024423/1. C.G. thanks Konstantin Bliokh for stimulating discussions prior to the commencement of this work.

\appendix

\section{Numerical solution of Schr\"{o}dinger equation}
\label{Appendix}

The time-dependent Schr\"{o}dinger equation can be solved numerically with high accuracy and efficiency by expanding the time-evolution operator in a series of Chebyshev polynomials.  In this appendix, we describe how this method can be applied to the two-dimensional Schr\"{o}dinger equation for an electron interacting with an external magnetic field.  For generality, we shall consider here a Hamiltonian of the form
\begin{eqnarray}
\label{scriptH}
\mathcal{H}(x,y) & = & S_2\left(\frac{\partial^2}{\partial x^2}+\frac{\partial^2}{\partial y^2}\right)+{\rm i} S_{1x}(x,y)\frac{\partial }{\partial x} \\ \nonumber
& & +{\rm i} S_{1y}(x,y)\frac{\partial}{\partial y}+S_0(x,y)
\end{eqnarray}
\cite{Dziubak2012}.
In the case of the Hamiltonian used in the main text, we would have $S_2 = -\hbar^2/(2m)$, $S_{1x} =-\hbar eBy/(2m)$, $S_{1y} =\hbar eBx/(2m)$ and $S_0 =e^2B^2(x^2+y^2)/(8m)$. The \emph{Mathematica} code we have used to perform these calculations has been made available online at \cite{Greenshields2015}.

Since the Hamiltonian \eqref{scriptH} is independent of time, we can write the solution of the  Schr\"{o}dinger equation as
\begin{equation}
\label{propagator}
\Psi(t+\Delta t)=\exp\left(-\frac{{\rm i}}{\hbar}\mathcal{H}\Delta t\right)\Psi(t).
\end{equation}
In order to evaluate this numerically, first we must represent the wavefunction, and the coefficients $S_2$ etc., on a two-dimensional grid.  If this grid covers an area $L_x \times L_y$, and contains $N_x\times N_y$ points, then the maximum spatial frequencies represented are $k_{x,{\rm max}}=\pi N_x/L_x$ and $k_{y,{\rm max}}=\pi N_y/L_y$.  The maximum and minimum values of energy represented on the grid are then
\begin{eqnarray}
E_{\rm max} & = & -S_2(k_{x,{\rm max}}^2+k_{y,{\rm max}}^2)+{\rm Max}(S_{1x})k_{x,{\rm max}} \\ \nonumber & & +{\rm Max}(S_{1y})k_{y,{\rm max}}+{\rm Max}(S_0)
\end{eqnarray}
and
\begin{equation}
E_{\rm min} = -\left[{\rm Max}(S_{1x})k_{x,{\rm max}}+{\rm Max}(S_{1y})k_{y,{\rm max}}\right]+{\rm Min}(S_0).
\end{equation}
Let us now introduce a new operator
\begin{equation}
\tilde{\mathcal{H}}=\frac{\mathcal{H}-b}{a},
\end{equation}
where $a=(E_{\rm max}-E_{\rm min})/2$ and $b=(E_{\rm max}+E_{\rm min})/2$.
This operator has eigenvalues represented on the grid which lie in the range $[-1,1]$.  We can then expand the time-evolution operator in a series of Chebyshev polynomials $T_q(\tilde{\mathcal{H}})$:
\begin{eqnarray}
\label{expansion}
\Psi(t+\Delta t) & = & \exp(-\frac{{\rm i}}{\hbar}b\Delta t)\exp(-\frac{{\rm i}}{\hbar}a\tilde{\mathcal{H}}\Delta t)\Psi(t) \nonumber \\
& \approx & \exp(-\frac{{\rm i}}{\hbar}b\Delta t)\sum_{q=0}^M\alpha_q(a\Delta t)T_q(\tilde{\mathcal{H}})\Psi(t).
\end{eqnarray}
Chebyshev polynomials are chosen as these minimise the error associated with truncating the expansion at a finite order $M$ \cite{TalEzer1984}.  The expansion coefficients are given by
\begin{equation}
\label{coefficients}
\alpha_q(a\Delta t) = \left\{
\begin{array}{lr}
(-{\rm i})^qJ_q(a\Delta t/\hbar), & q=0\\
2(-{\rm i})^qJ_q(a\Delta t/\hbar), & q\neq 0
\end{array}
\right.
\end{equation}
where $J_q$ is a Bessel function.  For $q>{\rm e}a\Delta t/(2\hbar)$, where e, in roman font, denotes the mathematical constant, the magnitudes of these coefficients decay exponentially with increasing $q$ \cite{Wang1998, Weisse2008}.  This means that the error due to truncating the series at order $M$, which can be estimated by $|\alpha_M(a\Delta t)|$, can be made arbitrarily small.  If we set 
\begin{equation}
M=\frac{\rm e}{2\hbar}a\Delta t+\delta,
\end{equation}
$\delta$ can be adjusted so that this error is less than machine precision.  The numerical error resulting from the Chebyshev expansion is then negligible.

In order to evaluate the individual terms in the expansion \eqref{expansion}, the action of the Chebyshev polynomial $T_q(\tilde{\mathcal{H}})$ on the initial wavefunction $\Psi(t)$ must be calculated.  Using the recurrence relation for the Chebyshev polynomials, we obtain
\begin{equation}
T_q(\tilde{\mathcal{H}})\Psi(t) = 2\tilde{\mathcal{H}}T_{q-1}(\tilde{\mathcal{H}})\Psi(t) - T_{q-2}(\tilde{\mathcal{H}})\Psi(t),
\end{equation}
for $q>0$, with the initial conditions $T_0(\tilde{\mathcal{H}})\Psi(t) = \Psi(t)$ and $T_1(\tilde{\mathcal{H}})\Psi(t) = \tilde{\mathcal{H}}\Psi(t)$.  The action of the Hamiltonian on the wavefunction can be efficiently calculated by evaluating the spatial derivatives in Fourier space \cite{Wang1998, Kosloff1983}.  That is,
\begin{eqnarray}
\mathcal{H}\Psi(t) & \approx& S_2{\rm FT}^{-1}[(-k_x^2-k_y^2){\rm FT}\Psi] +{\rm i} S_{1x}{\rm FT}^{-1}(ik_x{\rm FT}\Psi) \nonumber \\
& & + {\rm i} S_{1y}{\rm FT}^{-1}(ik_y{\rm FT}\Psi)+S_0\Psi,
\end{eqnarray}
where $k_x$, $k_y$ are the coordinates in Fourier space and FT denotes a discrete Fourier transform and ${\rm FT}^{-1}$ the corresponding inverse transform.


\end{document}